# X3D in Urban Planning - Savannah in 3D

Faith-Anne L. Kocadag and Felix G. Hamza-Lup
Armstrong State University
11935 Abercorn Street
Savannah, GA 31419

fk4687@stu.armstrong.edu, Felix.Hamza-Lup@armstrong.edu

## ABSTRACT

Urban planning often raises complex issues that are difficult to visualize and challenging to communicate. The increasing availability of 3D modeling standards has provided the opportunity for many developers, engineers, designers, planners, investors, and government officials to effectively collaborate to bring projects to fruition. Because of its real-time interactivity and widespread web-based content players, X3D proves to be a good choice for developing and visualizing 3D city content on the Web for planning purposes.

Passenger rail is a viable and cost-effective transportation solution in many areas, especially in view of rising energy costs. The Savannah in 3D (or S3D) project is a multimedia tool for a feasibility study designed to bring passenger rail to Savannah; thereby opening up the historic, tourist-friendly city to a wider audience. The paper outlines the development process of an interactive 3D train model as it journeys from Atlanta to Savannah, Georgia - focusing on user interactivity and scene immersion to supplement the city and transportation planning agenda.

## Categories and Subject Descriptors
H.5.1 [**Information Interfaces and Presentation**]: Multimedia Information Systems – *animations*.

## General Terms
Design

## 1. INTRODUCTION
Novel visualization tools and web-based programming standards like Extensible 3D (or X3D) have given developers the opportunity to stretch some artistic muscle while creating interesting and original content, potentially allowing even novice users to "play" on the field of 3D visualization. Savannah in 3D (S3D) is a project that shows the power of this lightweight open source standard as it pertains to city and transportation planning projects.

X3D and other 3D standards were employed in the past in transportation and urban planning. Yusuf, et al., improved the efficiency of simulation by combining X3D with artificial intelligence to create urban model options of varying accessibility [1]. Mao et al. [2], with an eye toward making city models publicly and widely available, integrated *CityTree* (a data structure for *CityGML*) with X3D to avoid overwhelming the user with mass data and economize load times [2, 3]. Verma, et al., modeled complex 3D buildings using combinations of simple parametric shapes based on data from aerial LIDAR (Light Detection and Ranging) [4], and Hermsmeyer, et al., demonstrated how 3D applications and topographic base data helped forecast and guide the brisk development of Kuwait City [5].

Creating content-rich 3D models requires cooperation among designers, developers, and experts in a myriad of fields. Sedlacek and Zara outlined many of the difficulties faced in the creation of a 3D reconstruction of the Langweil model of Prague, including the overabundance of data, repetitive textures that confused the algorithms, occluded geometry, and hardware limitations [6]. Even in the controlled environment of the City of Prague museum, constructing an accurate and useful model proved challenging.

The Savannah in 3D (S3D) project was conceived with the intention of providing a different viewpoint to the South by Southeast Rail feasibility study. The South by Southeast project is many-faceted and is in the midst of market research to "determine the demand for passenger rail … for business and tourism" [7]. Because of the intended scope of this project, X3D — in light of its portability, accessibility, and wide compatibility with web browsers — became the standard of choice. The 3D simulation provides an invaluable representation of what to expect if a passenger rail system is re-introduced in the Savannah area. The project investigates and projects a possible route between Atlanta and Savannah, Georgia, taking advantage, behind the scenes, of existing rail lines.

The paper outlines the scene creation process and demonstrates how S3D was designed to be engaging and user-friendly, thus providing the narrative for a memorable user experience. Further, we discuss the challenges encountered during implementation and discuss the future of such projects.

## 2. THE USER EXPERIENCE
### 2.1 The Georgia Scene
The Georgia scene provides a good introduction to the S3D project and can provide the South by Southeast organization a valuable tool for representing their vision for the future of rail and the advancement of public transportation in Georgia. Over time, the map could convey more information to the user, such as station information, landmarks, weather, or historical data. Also, augmentation of Google Earth™, with our X3D content is possible.

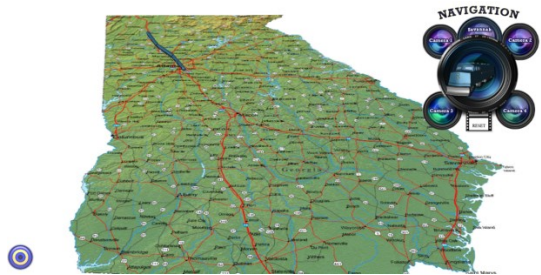

**Figure 1. Georgia Overhead View**

The simulation begins with the overhead view of a map of Georgia, as in Figure 1. A navigation menu (Figure 2) appears on the top right corner, allowing the user to choose from several viewpoints, trigger the train animation, and reset the train to its original position. Each viewpoint on the Georgia map is synchronized with the train animation, orienting the viewer with the most interesting and dynamic angles to view the train as it travels from Atlanta to Savannah. Once the train enters Savannah, the user has the option of navigating to the Savannah scene for a closer and detailed look. This paradigm improves the user experience by providing additional scene details like stylistically appropriate building facades layered with a jazzy, downtown, riverfront soundscape.

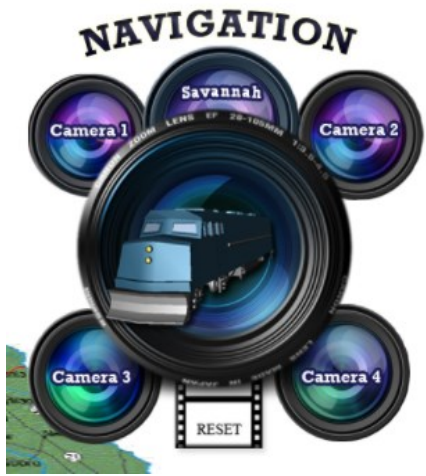

**Figure 2. The Navigation Menu**

## 2.2 The Savannah Scene

The Savannah scene is fundamentally different from the Georgia scene. While the focus on the Georgia scene is the journey, the Savannah scene is more concerned about the destination. There is freedom to "walk" around downtown Savannah. This interactivity is key to imparting the user with a constructive experience. Merely augmenting a simple map with a third dimension is insufficient. In detailing the strengths and weaknesses of X3D in software visualization, Anslow noted that additional control of 3D data is imperative [8].

While projects like Google Maps™ and Google Earth™ provide maps and occasionally 3D content, S3D has more of an eye toward local atmosphere and can be augmented with neighborhood tips, historical information, or can be modified to test and model statistics and progress from a civil engineering standpoint. While S3D is more stylized in its presentation and scaling, the overall experience is made more immersive through different lighting effects, sound, and animation events.

Moreover, while some may have trouble gleaning important information from a traditional map, S3D allows for a more guided experience, much akin to the way an Infographic passes on data more quickly and effectively than simple text. "Public interest and understanding can be raised by using both graphic and image files that show the relationship between the proposed project and individual properties, neighborhoods, local landmarks, community services, and other features" [9]. The goals of the South by Southeast project can be more clearly met through increasing interactivity, user engagement and allowing users to gain knowledge through exploration and insight.

The Savannah scene also begins with an overhead view (Figure 3) and a navigation menu in the top right corner. This time there are two trains to choose from, one incoming and one outgoing. Engine level viewpoints are available from both trains, allowing the user to view the city as the incoming train arrives at the Savannah station, as well as from the outgoing train as it exits the city. The user has the option of exploring Savannah independently, walking down historic streets and through the city square markets.

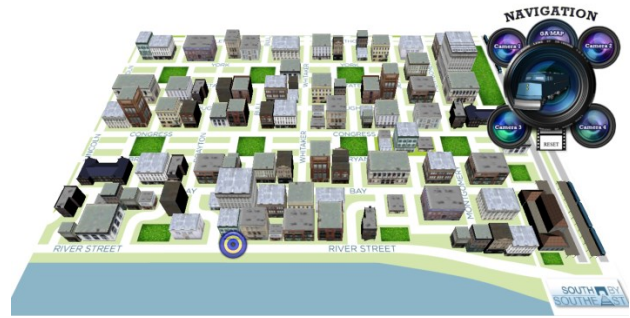

**Figure 3. Savannah Overhead View**

## 3. SCENE IMPLEMENTATION
### 3.1 Structure and Interactivity

The Georgia map scene and the Savannah scene are loaded simultaneously. The Savannah scene takes advantage of X3D's Inline node and is placed 500 meters "below" the Georgia scene (as illustrated in Figure 4). The Level of Detail (LOD) nodes ensure that neither scene is visible from the other.

On standard computer setups, the Georgia scene is drawn instantaneously and, on average, the music starts and the animation triggers are ready after 3.5 seconds. Both scenes together contain or call on 629 shapes, 74 *ImageTextures*, and one .WAV audio clip in five X3D scene-graphs (and an optional sixth).

The Georgia scene only requires two of those *ImageTextures* and two Inline notes for train geometry. Meanwhile, no disruptions occur as the more resource-heavy Savannah scene loads in the background. Once the user is ready to navigate to the Savannah scene (via the Navigation menu), the Savannah scene has fully downloaded and can be drawn instantly. Structuring the two scenes in this way allows for seamless transitions and mitigates any delays.

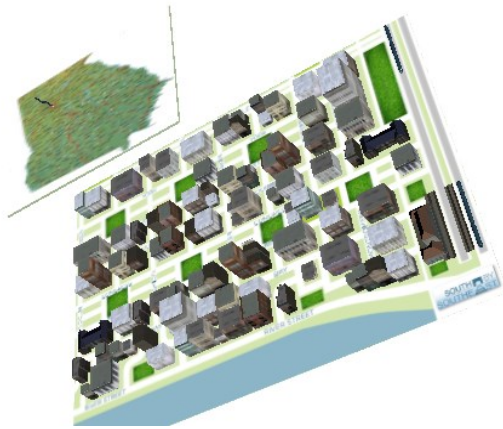

**Figure 4. Enhanced View of Scene Structure
with LODs Removed**

## 3.2 Facilitating Exploration

One of the biggest challenges with X3D is making the scene easily navigable. On most X3D players, the Page-Up and Page-Down keys can be utilized to scroll sequentially forward or backward through the bind-able viewpoint nodes in the scene-graph. Unfortunately, inexperienced computer users often ignore these keyboard shortcuts, or simply don't know where they are on the keyboard. Alternatively, users can right-click in a scene to choose different viewpoints and read their descriptions from a menu. This too, often becomes clumsy and inefficient. To simplify the navigation process, and to allow users to take full advantage of all available viewpoints, the Navigation Menu was introduced.

The Navigation Menu is an intuitive tool for navigating through different viewpoints and between the scenes. Proximity sensors in each of the two main scenes determine whether the menus are drawn and detect any positional/orientation changes of the viewer. Navigation Menus, in effect, move simultaneously with the user, so they appear stationary within each scene.

## 3.3 Viewpoints

Much in the same way digital video discs (DVDs) provide extra content through different camera angles, this S3D project provides different viewpoints from which the user can choose. Instead of scrolling through consecutive bind-able viewpoint nodes with the Page-Up and Page-Down keys, the Navigation Menu allows the user to choose a viewpoint at will. The X3D viewer automatically animates smooth transitions between the viewpoints for enjoyable user experience.

Camera 1 in the Georgia scene is the Overhead View. This is the only static viewpoint in this scene, and from this position, one can easily review the train route and map of Georgia. To see the train from every angle, the user should choose Camera 2. The camera semi-circles the train through the combination of an Orientation Interpolator and a Position Interpolator. The position and rotation of this camera can be visualized by "commenting-in" the shape node within the Georgia Moving Camera Transform (Figure 5).

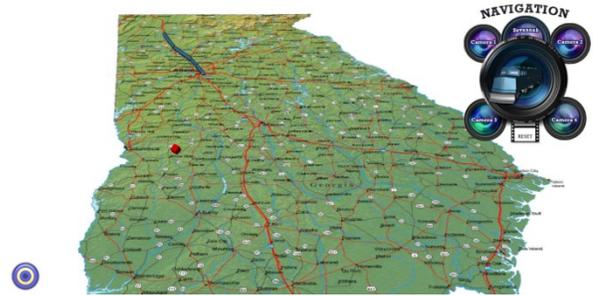

**Figure 5. A Red Cube Marks Moving Camera Position**

Choosing Camera 3 (Engine Level) gives the user a first-person view, allowing the user to "ride" the train firsthand. The Georgia Ground Level viewpoint (Camera 4) stations the user at ground level. Through the implementation of an Orientation Interpolator, the user maintains position and turns to watch the train as it passes.

In the Savannah scene, the Overhead View (Camera 1) provides a nice survey of a portion of downtown Savannah. The Savannah Ground Level View (Figure 6) drops the user on River Street and leaves them free to explore the city, while the train station is highlighted at the Savannah Train Station (Camera 2, Figure 7).

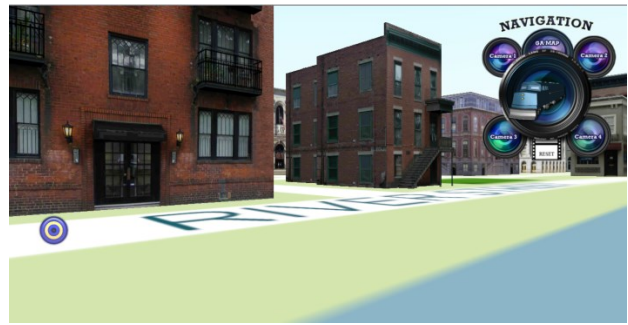

**Figure 6. Savannah Ground Level View**

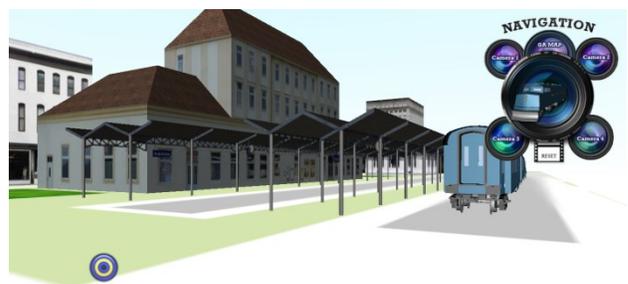

**Figure 7. Savannah Train Station View**

From Cameras 3 and 4, users are treated to the most visceral experiences. These two engine level viewpoints immerse the user in the ride and allow the Savannah cityscape to engage the user as the Incoming Train (Figure 8) and the Outgoing Train (Figure 9) make their way.

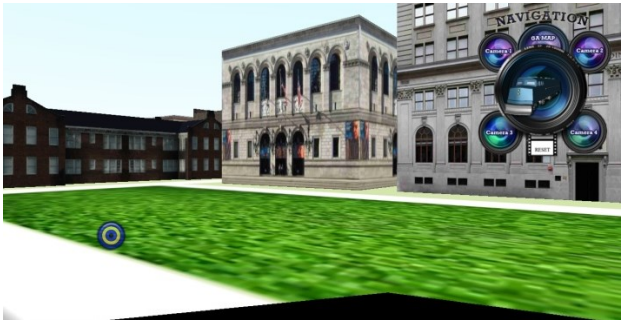

**Figure 8. Savannah Incoming Train View**

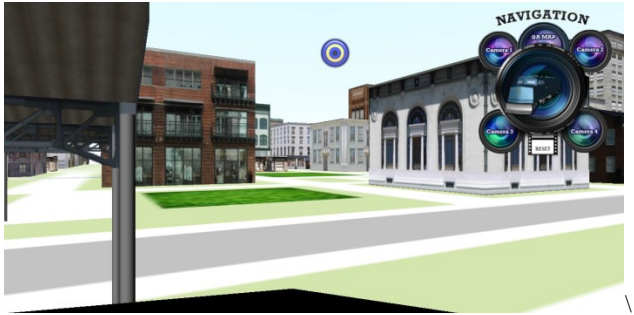

**Figure 9. Savannah Outgoing Train View**

The time a user takes to explore the train from different angles and "walk" around the city of Savannah also affords the developers an opportunity to inject meaningful data into the simulations. Users can learn interesting facts about the benefits of a passenger rail system in visual, not simply academic, terms.

A table of available Viewpoints is provided below (Table 1).

**Table 1. Available Bindable Viewpoints**

| Bindable Viewpoint | Static vs. Animated |
|---|---|
| Georgia Overhead | Static |
| Georgia Ground Level | Animated |
| Georgia Engine Level | Animated |
| Georgia Moving Camera | Animated |
| Savannah Overhead | Static |
| Savannah Ground Level | Static |
| Savannah Train Station | Static |
| Savannah Incoming Train | Animated |
| Savannah Outgoing Train | Animated |

## 3.4 Animation Events

There are several types of animation events in S3D: Position Interpolators, Orientation Interpolators, and Color Interpolators, most notably. In several cases, Transform nodes for Position Interpolators were placed within Transform nodes for Orientation Interpolators (and vice-versa) to provide a dynamic effect.

An example of a Position Interpolator at work is with the train animation in the Georgia scene. All non-rotational changes in the position of the train are controlled by a single Position Interpolator. Arranged within this interpolator are three nested Orientation Interpolators, each charged with handling the movement of a subsection of the train. Structuring the animation interpolators in this way and synchronizing them to the same *TimeSensor* keeps the train on its tracks and creates a fish-tail effect, where the cars appeared to rotate independently while each leading end remained fixed relative to the car before it.

Simpler Positional Interpolators were employed for the trains in the Savannah scene. These linear transformations slow the Incoming Train as it nears the station and speeds up the Outgoing Train upon departure. The individual cars were grouped similarly to the train in the Georgia scene, making the animations easily adaptable when train routes are changed and updated.

The Georgia train animation and all of the animated viewpoints in the Georgia scene are triggered simultaneously in two ways: by clicking on the train itself, or by clicking the picture of the train in the Navigation Menu. Both have the same effect, as both are *TouchSensors* connected to the same *TimeSensor* controlling all of the non-Spotlight animations of the scene.

Another animation effect, a Color Interpolator, was employed in conjunction with the Spotlight node in the Georgia scene to simulate the changes in sunlight throughout the day. This lighting effect is detailed in the next section.

The X3D browser uses Position, Orientation, Scalar, and Color Interpolators as a map for calculating transitions/animations between *keyValues*. The fewest possible number of *keyValues* were used to create desired effects while attempting to keep changes in direction and position as smooth and seamless as possible. Higher numbers of *keyValues* in positional interpolators may have kept the affected geometry on a tighter path, but trial-and-error proved that these changes were often more abrupt or "jerky" than desired.

## 3.5 Background, Lighting, and Sound

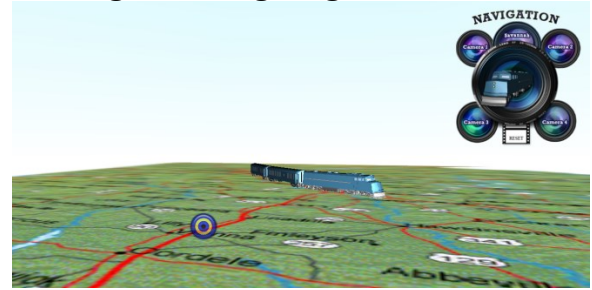

**Figure 10. Train under Spotlight**

Geometry and animation events alone were not sufficient in creating a full, immersive environment. Backgrounds, lighting, and sound also played vital roles. The Background node of the scene simulates a blue sky which smoothly transitioned to the white background below. The blue sky was important for orienting the user, letting them know which end is up, and giving the illusion of a sunny day. While a clear sky was appropriate, as Savannah averages 216 sunny days each year, modeling other weather conditions would be very useful from a city or transportation planning standpoint. Fog nodes already exist in the X3D library, and others like Murphy [10] have developed algorithms for rendering 3D clouds in the X3D schema for meteorological purposes. Having a web browser download current weather conditions to input into the scene would be a worthwhile extension of this project.

To make the scene more realistic, a Spotlight node was added to the center of the Georgia scene to illuminate the train as it passes. The Spotlight, representing the sun, was positioned toward the center of the route so the user could appreciate the changing specular color of the train on its journey. The Color Interpolator transitions the color of the light from yellow to white to yellow in 12 second cycles, mimicking the changes in sunlight throughout the day.

From the source code, an advanced user could "uncomment" the inline node of a textured block (WhiteRectangleBackdrop.x3d) that illuminates the features of the Spotlight, including position, color, intensity, and attenuation. Figure 11 illustrates an example.

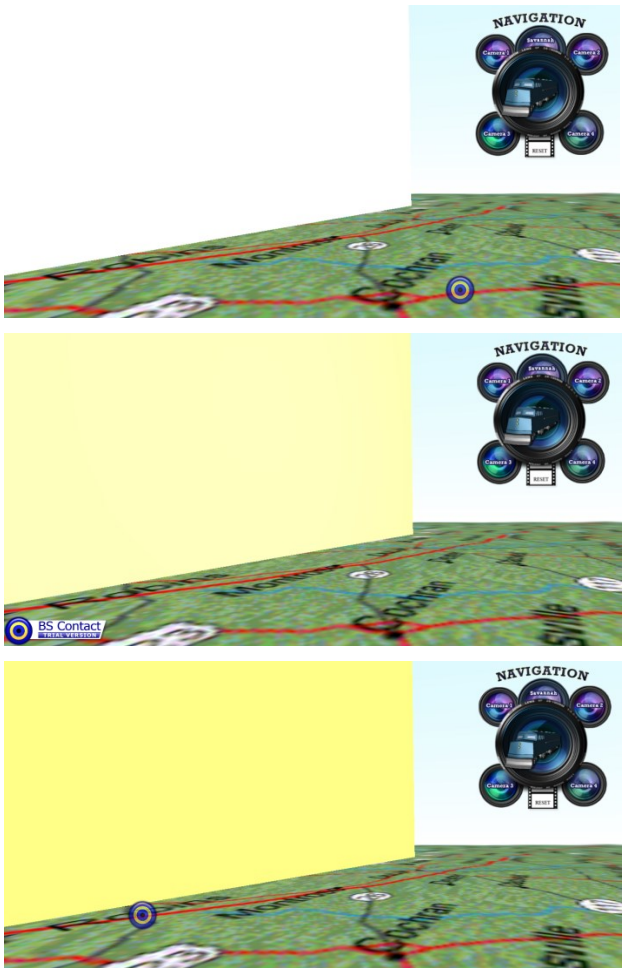

**Figure 11. "*WhiteRectangleBackdrop*" illuminates with the Changing Spotlight**

Another layer of realism is added to S3D through the use of an AudioClip node. The short ambient loop of train, waterfront, and city sounds added style and texture to elevate the user experience.

### 3.6 Optimization

Keeping the user immersed and engaged with the scene was of the utmost importance, so it was crucial to preclude slow downloads and long loading times. In addition to compartmentalizing the Georgia and Savannah scenes through LODs, StaticGroup nodes were implemented wherever possible, and the scene (including all inline nodes) were converted from traditional X3D (.x3d) to X3D binary (.x3db) encoding. Placing static objects in StaticGroups freed memory resources at runtime. Binary compression reduced file sizes an average with 68% (see Table 2 for details), suggesting a noticeable improvement in download time for the user. Stocker and Schickel [11] asserted that binary compression more effectively and economically represented meshes, and increased loading speed by a factor of two to four.

**Table 2. X3D vs. X3D with Binary Compression**

| File Size (in bytes) | .x3d | .x3db | % Reduction |
|---|---|---|---|
| Georgia Scene | 11,791 | 4,808 | 59.22 |
| Savannah Scene | 98,078 | 37,494 | 61.77 |
| Train Station | 54,717 | 25,481 | 53.43 |
| Train Engine | 502,209 | 63,766 | 87.30 |
| Train Car | 391,858 | 73,575 | 81.22 |
| Average Reduction | | | 68.59 |

## 4. CONCLUSIONS AND FUTURE WORK

The Savannah in 3D [12] project provided an immersive 3D experience for a proposed passenger rail system, one that can be easily improved as more findings of the South by Southeast project studies emerge. Usability was a prime concern, so the focus was placed on software accessibility, intuitive and efficient navigation design, the economy of resources (optimization), and providing a meaningful narrative that South by Southeast could utilize to further their objectives.

In the short term, once an exact location is formalized, this project would be able to simulate detailed, accurate representations of the area around the Savannah station. Additional expansion can come with the modeling of other notable cities along the train route, including Atlanta and Macon, Georgia. Further, down the line, as the 3D technology continues to permeate to standard devices, the project could be augmented as a marketing tool to provide a comprehensive 3D experience to possible commuters and tourists.

Other teams, like Zuffo, et al., have integrated historical data into X3D scene-graphs; creating historical reconstructions [13]. Given the rich history of Savannah and the expansiveness of the downtown historic district, such a period reconstruction would prove invaluable to historians, students, and tourists.

Extensible 3D has proven a robust instrument for this project. Many of its different nodes were used in the development of S3D, and oftentimes in unexpected ways. X3D, and 3D modeling in general is an invaluable tool for authors to learn to express their visions, create, propagate, and possibly even master complex geometry.

## 5. ACKNOWLEDGMENTS

We would like to acknowledge Eckard "Andy" Cabistan for his support in this project. Thanks also to Tobias Merk and Jason Combs for allowing us to adapt and convert several of their 3D models to augment this project.